\RequirePackage{snapshot}
\documentclass[aps,pra,twocolumn,superscriptaddress,nofootinbib,showpacs]{revtex4-1}

\usepackage[english]{babel}
\usepackage{amsmath}
\usepackage{amsfonts}
\usepackage{amssymb}
\usepackage{amsthm}
\usepackage{graphicx}
\usepackage{braket}

\begin{document}

\title{Coherent destruction of tunneling in graphene irradiated by elliptically polarized lasers}
\author{Denis Gagnon}
\email{denis.gagnon@uwaterloo.ca}
\author{Fran\c{c}ois Fillion-Gourdeau}
\affiliation{Universit\'e du Qu\'ebec, INRS--\'Energie, Mat\'eriaux et T\'el\'ecommunications, Varennes, Qu\'ebec, Canada, J3X 1S2}
\affiliation{Institute for Quantum Computing, University of Waterloo, Waterloo, Ontario, Canada, N2L 3G1}
\author{Joey Dumont}
\affiliation{Universit\'e du Qu\'ebec, INRS--\'Energie, Mat\'eriaux et T\'el\'ecommunications, Varennes, Qu\'ebec, Canada, J3X 1S2}
\author{Catherine Lefebvre}
\author{Steve MacLean}
\email{steve.maclean@uwaterloo.ca}
\affiliation{Universit\'e du Qu\'ebec, INRS--\'Energie, Mat\'eriaux et T\'el\'ecommunications, Varennes, Qu\'ebec, Canada, J3X 1S2}
\affiliation{Institute for Quantum Computing, University of Waterloo, Waterloo, Ontario, Canada, N2L 3G1}
\date{\today}

\begin{abstract}
	
Photo-induced transition probabilities in graphene are studied theoretically from the viewpoint of Floquet theory.
Conduction band populations are computed for a strongly, periodically driven graphene sheet under linear, circular, and elliptic polarization.
Features of the momentum spectrum of excited quasi-particles can be directly related to the avoided crossing of the Floquet quasi-energy levels.
In particular, the impact of the ellipticity and the strength of the laser excitation on the avoided crossing structure -- and on the resulting transition probabilities -- is studied.
It is shown that the ellipticity provides an additional control parameter over the phenomenon of coherent destruction of tunneling in graphene, allowing one to selectively suppress multiphoton resonances.

\end{abstract}

\maketitle

\section{Introduction}

Graphene is a remarkable 2D material with properties particularly suited to the development of electronic and optoelectronic devices \cite{CastroNeto2009, Engel2012}, for instance ballistic transistors \cite{Geim2007} and photo-detectors \cite{Schall2014}.
Besides this promise of applications, the Dirac band structure of graphene has motivated several studies of the fundamentals of light-matter interactions.
In particular, a celebrated property of pristine graphene is its transformation in a Floquet topological insulator (FTI) upon irradiation with a circularly polarized pulse \cite{Cayssol2013}.
In regular TIs, the band structure is modified by spin-orbit interactions and edge states are protected by time-reversal symmetry \cite{Bansil2016}.
However, in the case of graphene, a circularly polarized field breaks time-reversal symmetry and opens a gap at the Dirac point, leading to a FTI state which cannot exist in the absence of incident radiation \cite{Usaj2014}.
This gives rise to Floquet edge states confined to the boundaries of graphene ribbons, states with definite chirality and robustness against disorder \cite{Perez-Piskunow2014}.

Another quantum optical phenomenon which has garnered some attention in recent work is the realization of Landau-Zener-St\"{u}ckelberg interference (LZSI) in graphene irradiated by a linearly polarized excitation.
LZSI has been generically described for superconducting qubits \cite{Shytov2003, Shevchenko20101}, Majorana qubits \cite{Wang2016}, and also in theoretical \cite{Kelardeh2016, Fillion-Gourdeau2016, Rodionov2016} and experimental \cite{Higuchi2016} works specific to graphene.
In both physical realizations (qubits and graphene), quantum interference stems from the periodic driving of the system through avoided crossings of its adiabatic energy levels \cite{Shevchenko20101}.
In the case of a circularly polarized excitation (which leads to the FTI state in graphene), LZSI is absent since the system no longer goes through the aforementioned avoided crossings. 

This article presents a theoretical study of the modification of Floquet quasi-energies and transition probabilities in graphene when an incident monochromatic laser is tuned from a linear polarization to an elliptic polarization.
From a fundamental standpoint, the photo-induced conduction band population calculated using Floquet eigenvectors is formally analogous to the production of electron-positron pairs from vacuum, as discussed in several recent articles \cite{Allor2008, Lewkowicz2011, Dora2010, Fillion-Gourdeau2015, Fillion-Gourdeau2016}.
This population could, in principle, be probed using angle-resolved photoemission spectroscopy (ARPES) on graphene \cite{Sentef2015, Kelardeh2016}.
The main goal of this article is to identify the various physical features of the Floquet momentum maps that appear or disappear as the ellipticity parameter is varied, for instance coherent destruction of tunneling (CDT) \cite{Grossmann1991, Grifoni1998, Son2009}.
Our results show that the ellipticity of the polarization provides a control parameter over multiphoton processes in graphene.
In particular, for some range of value of the ellipticity of the incident field and for some momentum states, electron-hole pair production can be suppressed for several given field strengths.

\section{Floquet formulation for graphene} 

The Floquet formulation is an appropriate tool for the study of strongly, periodically driven quantum systems.
It allows one to recast a time-dependent problem into a time-independent one, thereby allowing the use of all tools specific to time-independent quantum mechanics, such as stationary perturbation theory \cite{Grifoni1998, PhysRevA.75.063414}.
Several articles have been concerned with the application of Floquet theory to account for various physical phenomena in graphene.
These include the previously mentioned emergence of the FTI state under circularly polarized illumination \cite{Cayssol2013,Usaj2014},
the prediction of ARPES results \cite{Sentef2015}, the motion of Dirac points in the tight-binding model \cite{Delplace2013, Rodriguez-Lopez2014}, the photo-voltaic Hall effect \cite{Oka2009}, the presence of bound states around adatoms \cite{Lovey2016}, and quantum interference in the case of linear polarization \cite{Fillion-Gourdeau2016, Rodionov2016}.
In this section, we review the application of the Floquet formalism to graphene, the ultimate goal being the computation of transition probabilities in momentum space for an elliptically polarized laser excitation.

\begin{figure}
	\includegraphics[width=0.4\textwidth]{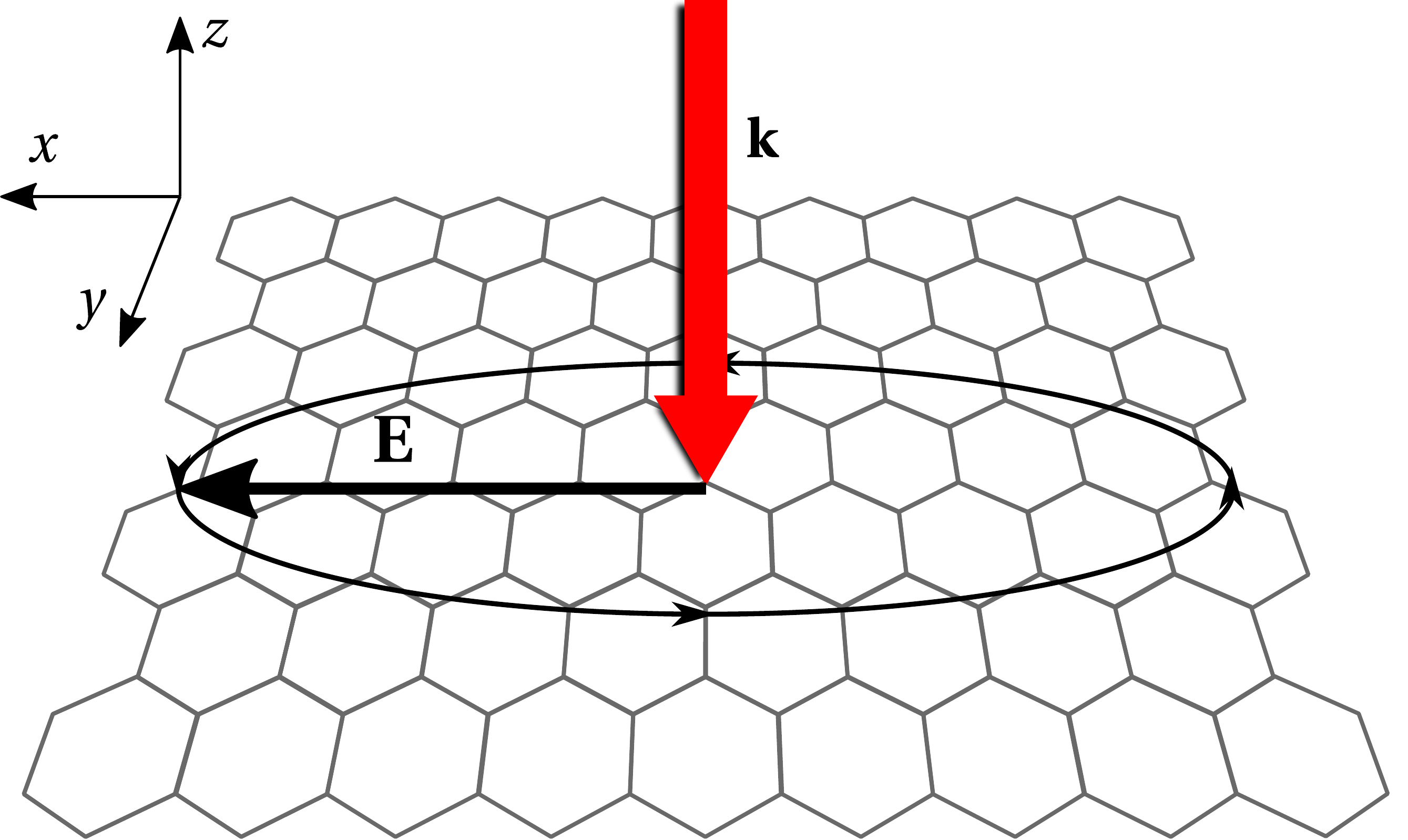}
	\caption{Graphene sheet irradiated by an elliptically polarized excitation.}
	\label{fig:graphene}
\end{figure}

Consider Dirac fermions in a graphene mono-layer in the presence of an elliptically polarized laser field, which is uniform in space and normally incident to the graphene plane (see Fig. \ref{fig:graphene}).
The fermion dynamics are governed by the following $(2 \times 2)$ low energy Hamiltonian (we use units such that $\hbar = 1$)
\begin{equation}\label{eq:graphene1}
H_{\xi}(t,\mathbf{p}) = \xi v_F \boldsymbol{\sigma} \cdot [\mathbf{p} + e \mathbf{A}(t) ],
\end{equation}
where $\xi = \pm 1$ is the valley pseudospin index, $v_F \simeq c /300$ is the Fermi velocity in graphene, $\boldsymbol{\sigma} = (\sigma_x, \sigma_y)$ is a vector of Pauli matrices representing the sublattice pseudospin, $\mathbf{p}$ is the fermion momentum around the $K^\xi$ points and $-e < 0$ is the electron charge.
The vector potential used to model the elliptically polarized excitation is given by
\begin{equation}\label{eq:vector_potential}
\begin{aligned}
A_x(t) & = A_0 \cos \omega t,  \\
A_y(t) & = \eta A_0 \sin \omega t,
\end{aligned}
\end{equation}
where $\eta$ is the ellipticity of the excitation and $A_0 \equiv E_0 / \omega$ is the maximum amplitude of the vector potential.
As shown in Fig. 1, this vector potential describes a field precessing around the $z$ axis in an elliptic pattern, with a semi-major axis parallel in the $x$ direction and a semi-minor axis in the $y$ direction.
A change of representation of the Dirac matrices can be performed via a unitary transformation (which does not impact physical observables) to obtain a Hamiltonian closer to the canonical form of a strongly driven two-level atom \cite{Fillion-Gourdeau2016}:
\begin{eqnarray}
\label{eq:unitary}
U_{r}\equiv e^{-i\sigma_{y}\frac{\pi}{4}} e^{-i\sigma_{x}\frac{\pi}{4}}.
\end{eqnarray}  
Combining this transformation with the following change of variables
\begin{subequations}
	\begin{align}
	\varepsilon_{i} & \equiv  -2\xi v_F p_{i}, \\
	A'_0 &\equiv   -2\xi v_F  e E_0/\omega,
	\end{align}
\end{subequations}
leads to
\begin{equation}\label{eq:two_level}
\begin{aligned}
H_{\xi}(t,\mathbf{p}) = -\frac{1}{2}
\begin{pmatrix}
\varepsilon_x + A'_0 \cos \omega t	& \varepsilon_y + \eta A'_0 \sin \omega t\\
 \varepsilon_y + \eta A'_0 \sin \omega t	& - \varepsilon_x - A'_0 \cos \omega t
\end{pmatrix}.
\end{aligned}
\end{equation}

One of the goals of this article is to explore the parametric range $0 \leq \eta \leq 1$ using Floquet theory, and to show the various physical effects that occur in the bulk as the system transitions to the FTI phase.
In the case of a linearly polarized excitation ($\eta = 0$), the Hamiltonian in Eq. \eqref{eq:two_level} is of the generic form describing a driven two-level system, or a particle moving in a double-well potential \cite{Grifoni1998}.
The transverse momentum $\varepsilon_y$ plays the role of a coupling strength between the two basis states, or two wells \cite{PhysRevA.75.063414, Shevchenko20101}.
Thus, other quantum systems described by this Hamiltonian include atoms in intense laser fields \cite{Shevchenko20101}, semiconductor superlattices \cite{Holthaus1992, Rotvig1995, Platero2004}, and superconducting qubits \cite{Oliver2005, Son2009}, where $\varepsilon_y$ is generally called ``tunnel splitting''.
The terminology ``transverse momentum'' is also used to describe $\varepsilon_y$, as it is proportional to the quasi-particle momentum in the direction perpendicular to the (linearly polarized) electric field.
In the case of a circularly polarized excitation ($\eta = 1$), this definition is ambiguous, but we shall retain the terminology ``transverse momentum'' in the elliptic case with the understanding that it is the momentum component perpendicular to the semi-major axis of the polarization ellipse.

The Fourier series of a periodic $2 \times 2$ Hamiltonian $H(t)$ reads (we use the notation of Son \emph{et al.} \cite{Son2009} in the following derivation)
\begin{equation}
H(t) = \sum_{n=-\infty}^{\infty} H^{[n]} e^{-i n \omega t},
\end{equation}
where $n$ is an integer, the Fourier index. The Floquet state nomenclature reads
\begin{equation}
\ket{\nu n} = \ket{\nu} \otimes \ket{n},
\end{equation}
where $\nu$ is the system index which can take two values, $\alpha$ and $\beta$, labeling negative and positive energy eigenstates of $\sigma_z$, respectively.
Switching to Fourier space and applying the Floquet theorem yields
\begin{equation}\label{eq:eigenvalue}
\sum_{\mu = \alpha, \beta} \sum_{m} \bra{\nu n} H_F \ket{\mu m} \left\langle \mu m | q_{\gamma l} \right\rangle = q_{\gamma l} \left\langle \nu n | q_{\gamma l} \right\rangle,
\end{equation}
where $q_{\gamma l}$ are the Floquet quasi-energies, $\ket{q_{\gamma l}}$ are the Floquet eigenvectors, and $H_F$ is the Floquet Hamiltonian.
This Floquet Hamiltonian has a block matrix structure given by
\begin{equation}
\bra{\mu n} H_F \ket{\nu m} = H^{[n-m]}_{\mu \nu} + n \omega \delta_{\mu \nu} \delta_{nm}.
\end{equation}
Starting from \eqref{eq:two_level}, one can write the three non-vanishing $2 \times 2$ Fourier components of the Floquet Hamiltonian:
\begin{subequations}
\begin{equation}
H^{[0]} = -\frac{1}{2}
\begin{pmatrix}
\varepsilon_x & \varepsilon_y \\
\varepsilon_y & - \varepsilon_x
\end{pmatrix},
\end{equation}
\begin{equation}
H^{[\pm 1]} = -\frac{A'_0}{4}
\begin{pmatrix}
1 & \pm i \eta \\
\pm i \eta  & - 1
\end{pmatrix}.
\end{equation}
\end{subequations}

The transition probability between field-free eigenstates $\ket{-}$ and $\ket{+}$ (which have negative and positive eigenenergies, respectively) can be written as a sum of $k$-photon transition probabilities \cite{Son2009}
\begin{equation}
\label{eq:probability}
\bar{P}_{\ket{-} \rightarrow \ket{+}} = \sum_k \sum_{\gamma l} \left| \left\langle +, k | q_{\gamma l} \right\rangle \left\langle q_{\gamma l} | -, 0 \right\rangle \right|^2,
\end{equation}
where
\begin{equation}
\ket{-,k} = \frac{\varepsilon_x + |\varepsilon|}{\mathcal{N}} \ket{\alpha k} + \frac{\varepsilon_y}{\mathcal{N}} \ket{\beta k}  ,
\end{equation}
\begin{equation}
\ket{+,k} = - \frac{\varepsilon_y}{\mathcal{N}} \ket{\alpha k} + \frac{\varepsilon_x + |\varepsilon|}{\mathcal{N}} \ket{\beta k}  ,
\end{equation}
\begin{equation}
|\varepsilon| \equiv  \sqrt{\varepsilon_x^2 + \varepsilon_y^2}, \qquad \mathcal{N} \equiv  \sqrt{ (\varepsilon_x + |\varepsilon|)^2 + \varepsilon_y^2}.
\end{equation}

There generally exists no closed form solution for the transition probability, except for a special case of linear polarization discussed below.
The momentum-integrated density of photo-excited pairs can then be calculated from the transition probability using the following definition \cite{Fillion-Gourdeau2015}
\begin{equation}
\bar{n} = \frac{1}{(2 \pi)^2} \int  \bar{P}_{\ket{-} \rightarrow \ket{+}} d^2 p.
\end{equation}

\section{Numerical results and discussion}

We now turn our attention to numerical calculations of the transition probability between the valence and conduction band of graphene, Eq. \eqref{eq:probability}, for several combinations of the coupled graphene-laser Hamiltonian parameters. 
These parameters are the normalized energy scales $\varepsilon_x / \omega$, $\varepsilon_y / \omega$ associated to the quasi-particle momentum, $A'_0/\omega$ associated to the applied field strength, and $\eta$.
The eigenvalue problem, Eq. \eqref{eq:eigenvalue}, can be solved numerically by constructing a truncated version of the Floquet Hamiltonian and using standard linear algebra routines.
The eigenvectors can then be used to compute the time-averaged transition probability between the valence and conduction band in graphene, Eq. \eqref{eq:probability}.
In all numerical calculations presented in this article, the Floquet Hamiltonian is truncated to 75 blocks to ensure a numerically converged solution, for a total matrix size of $302 \times 302$.

\subsection{Linear and circular polarization}

\begin{figure}
	\includegraphics[width=0.5\textwidth]{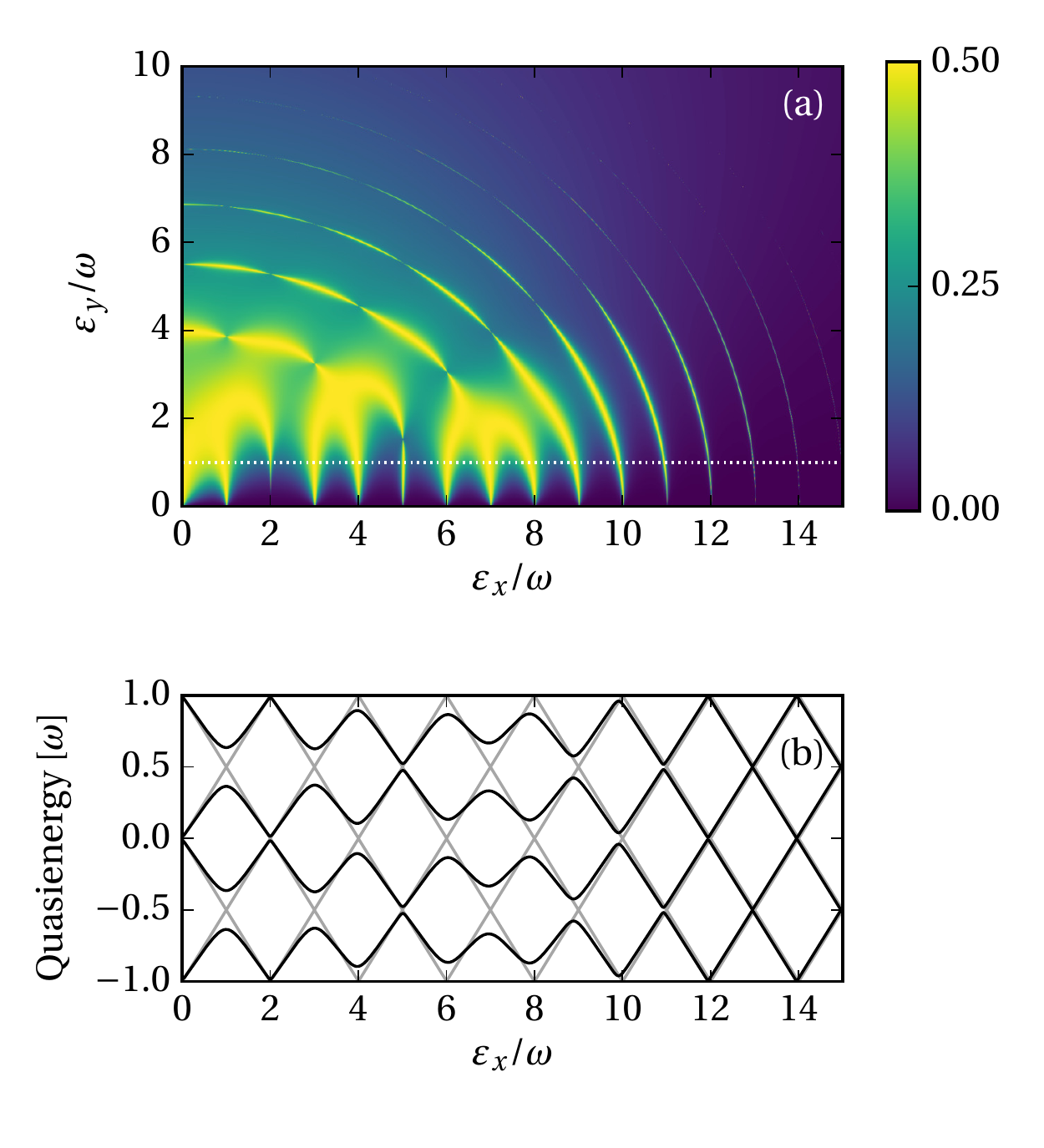}
	\caption{(a) Transition probability between field-free eigenstates $\ket{-}$ and $\ket{+}$ computed via Floquet theory.
		The field is linearly polarized ($\eta=0$), with $A'_0/\omega = 8.413$. 
		The momentum pattern is symmetric with respect to $\varepsilon_x / \omega = 0$ and $\varepsilon_y / \omega = 0$.
		(b) Evolution of the Floquet quasi-energies for $\varepsilon_y / \omega = 1$, indicated by a white line on the top figure.
		The relation between the momentum space resonances' widths and the quasi-energy separation is apparent. 
		For comparison, grey lines indicate the quasi-energies for $\varepsilon_y / \omega = 0$.}
	\label{fig:linear}
\end{figure}

Consider first the paradigmatic case of a linearly polarized field incident on the graphene layer ($\eta=0$).
For the purposes of this article, we use a normalized field strength $A'_0/\omega = 8.413$, which corresponds for example to a field of strength $E_{0} = 1.0 \times 10^{7}$ V/m oscillating at a frequency $f = \omega / 2 \pi = 10.0$ THz.
The resulting momentum space pattern is shown in Fig. \ref{fig:linear}a.
The appearance of multiphoton resonant fringes in the momentum space pattern of transition probabilities is directly related to the Floquet band diagram, as shown in Fig. \ref{fig:linear}b.
The location of a given peak usually corresponds to an avoided crossing between the Floquet quasi-energies, either of the same Fourier index or block (for even $k$-photon resonances) or of adjacent blocks (for odd $k$-photon resonances).
The quasi-energy gap is directly related to the resonance width: the closer the eigenstates anti-cross, the narrower the resonance.
In the extreme case of zero gap, the resonances actually disappear, as can be seen for instance at zero transverse momentum $\varepsilon_y / \omega = 0$.
Note that the maximal value of the transition probability is $1/2$ because it constitutes a time-average of oscillations between 0 and 1.
The corresponding momentum integrated density for $f = 10$ THz is $\bar{n} \simeq 2.5 \times 10^{11}$ cm$^{-2}$.

At small values of the transverse momentum, the presence of Lorentzian shaped multiphoton resonances is manifest.
This can be interpreted using a closed-form solution for the transition probability derived in Refs. \cite{Oliver2005, Son2009}.
In the limit $\varepsilon_y^2 \ll \varepsilon_x^2, |A'_0\omega|$, and considering $\eta = 0$, the field-free eigenstates can be approximated by those of $\sigma_z$, that is $\ket{\alpha}$ and $\ket{\beta}$.
Starting from Eq. \eqref{eq:probability}, a leading order perturbation treatment leads to the following analytic formula for the transition probability 
\begin{equation}
\label{eq:probability2}
\bar{P}_{\ket{\alpha} \rightarrow \ket{\beta}} = \sum_k \frac{1}{2}\frac{[\varepsilon_y J_k(A'_0/\omega)]^2}{[\varepsilon_y J_k(A'_0/\omega)]^2 + [k \omega - \varepsilon_x]^2 },
\end{equation}
where $J_k$ is the Bessel function of the first kind.
In other terms, for a linearly polarized laser excitation and a small transverse momentum $\varepsilon_y$, the time-averaged transition probability can be expressed as the superposition of Lorentzian $k$-photon resonances.
In this limit, the probability assumes its maximal value of $1/2$ at resonant values of $\varepsilon_x = k \omega$, and is suppressed at $\varepsilon_y = 0$ (zero transverse momentum).
The $k$-photon transition probability also tends to zero when $A'_0/\omega$ is equal to the $n$th zero $j_{k,n}$ of the Bessel function $J_k$, a situation which corresponds to CDT \cite{Grifoni1998, Son2009}.
This explains why the $k =2$ and $k=5$ multiphoton peaks are very narrow in Fig. \ref{fig:linear}a, since $A'_0/\omega = 8.413$ approaches $j_{2,2} = 8.41724$ and $j_{5,1} = 8.77148$.
The corresponding dressed quasi-enery gaps are accordingly small, as seen in Fig. \ref{fig:linear}b.

The momentum space patterns computed in the Floquet formulation (Fig. \ref{fig:linear}a) agree well with those obtained from solutions of the time-dependent Dirac equation when averaged over fast oscillations, as reported in an earlier contribution \cite{Fillion-Gourdeau2016}.
These time-dependent results can in turn be interpreted in terms of LZSI.
Pair production (or transition probability) peaks correspond to constructive interference between the different quantum pathways after an integer number of applied field cycles, and dips inversely correspond to destructive interference, equivalent to CDT \cite{Kelardeh2016}.
In fact, the small transverse momentum result, Eq. \eqref{eq:probability2}, can be obtained either via Floquet theory or using the adiabatic impulse model, the latter of which is commonly used to describe LZSI \cite{Shevchenko20101}.

\begin{figure}
	\includegraphics[width=0.4\textwidth]{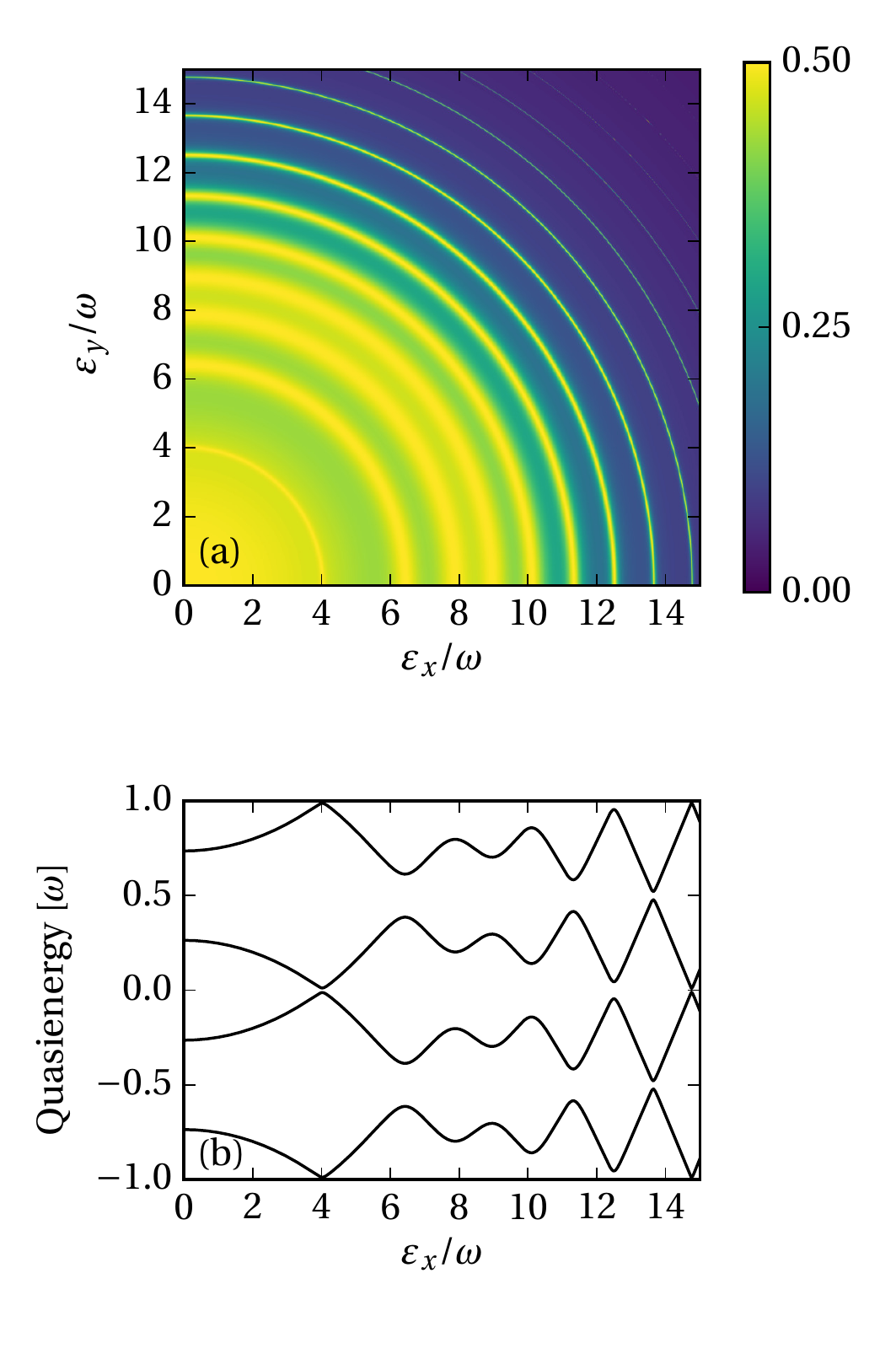}
	\caption{(a) Transition probability between field-free eigenstates $\ket{-}$ and $\ket{+}$ computed via Floquet theory.
		The field is circularly polarized with ($\eta=1$), with $A'_0/\omega = 8.413$. 
		The probability exhibits circular symmetry in momentum space.
		(b) Evolution of the Floquet quasi-energies for $\varepsilon_y / \omega = 0$.}
	\label{fig:circular}	
\end{figure}

At the other end of the spectrum, the transition probabilities for a circular polarization ($\eta=1$) are shown in Fig. \ref{fig:circular}a.
The momentum space pattern is now rotationally symmetric, and the circularly polarized excitation opens a gap in the quasi-energies at the Dirac point at small values of $|\varepsilon|$ (see Fig. \ref{fig:circular}b).
This photo-induced gap is consistent with the description of the quantum system in terms of a FTI.
Resonant multiphoton rings appear at fixed values of $|\varepsilon|$, the width of which is again directly related to the quasi-energy gap.
For large values of $|\varepsilon|$, the rings become sharp and distinct, as they do in the case of linear polarization ($\eta=0$).
This can be explained by the fact that the magnitude of the diagonal blocks of the Floquet Hamiltonian $H^{[0]}$ becomes much larger than the magnitude of the field-dependent off-diagonal blocks $H^{[\pm 1]}$.
In that case, the quasi-energy dispersion relation approaches that of undressed, or diabatic, eigenstates (c.f. Figs. \ref{fig:linear}b and \ref{fig:circular}b for large $\varepsilon_x/\omega$).
The corresponding momentum integrated density for $f = 10$ THz is $\bar{n} \simeq 6.3 \times 10^{11}$ cm$^{-2}$.
In the case of $0 < \eta < 1$ (discussed in the next subsection), we find that the momentum integrated density typically falls between this value and that of the linearly polarized case.

\begin{figure}
	\includegraphics[width=0.5\textwidth]{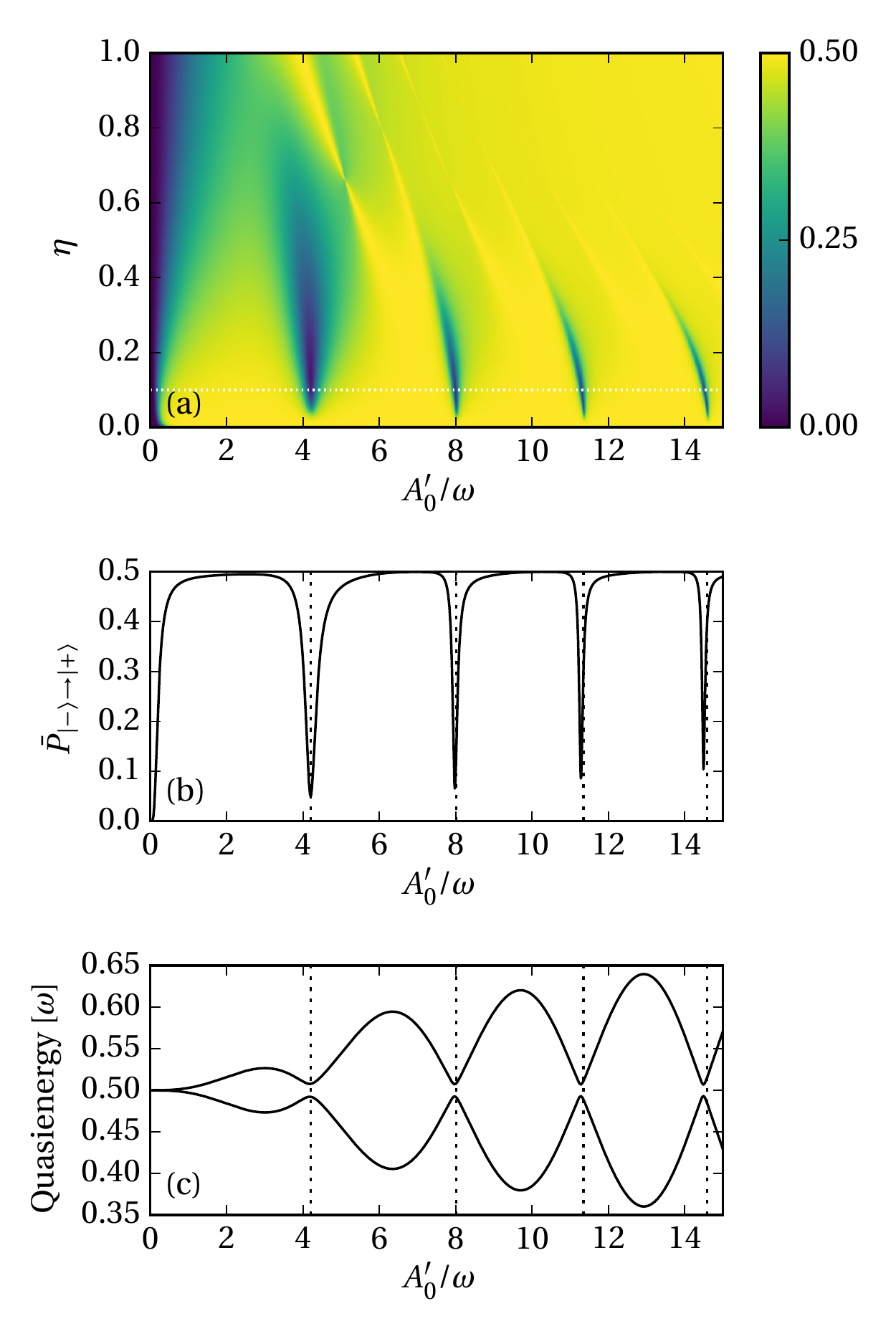}
	\caption{(a) Evolution of the 
		transition probability between field-free eigenstates $\ket{-}$ and $\ket{+}$ computed via Floquet theory as a function 
		of the field strength and ellipticity.
		This parameter sweep is computed for a resonant value of $\varepsilon_x / \omega = 3$ and $\varepsilon_y / \omega = 0.01$.
		(b) Evolution of the transition probability for $\eta=0.1$, indicated by a white line on the top figure.
		Dips in the probability value are noticeable.
		(c) Evolution of the Floquet quasi-energies for $\eta=0.1$.
		Black dashed lines indicate the expected position of the narrow avoided crossings (see Appendix).}
	\label{fig:emap}
\end{figure}

\subsection{Elliptic polarization}

Next, we study the effect of the ellipticity parameter of the incident field, $\eta$, on the Floquet transition probability with an emphasis on the phenomenon of CDT \cite{Grossmann1991}.
The condition for the occurrence of this phenomenon is a very close encounter of two quasi-energy levels, or in other words a nearly vanishing effective tunnel splitting.
In that case, quantum tunneling is brought to an almost complete standstill and the transition probability is suppressed.
This is due to the fact that the Rabi frequency of population oscillations is directly proportional to the effective tunnel splitting (or quasi-energy gap), and thus approaches zero \cite{Grossmann1991, Grifoni1998}.
Another way of interpreting this result is that as the quasi-energy gap reduces, the resonances narrow to the point of disappearing completely, as seen in the linearly polarized case described above for $\varepsilon_y/\omega \rightarrow 0$ (c.f. Fig. \ref{fig:linear}).
The CDT phenomenon has also been described in semiconductor nanostructures, which are governed by a similar Hamiltonian \cite{Holthaus1992, Rotvig1995, Platero2004}.
In a AC driven semiconductor superlattice, for example, CDT can occur for values of the Bloch frequency which minimize the effective tunnel splitting, an effect also known as ``miniband collapse'' \cite{Rotvig1995}.

Let us now consider a multiphoton resonance situation, i.e. $\varepsilon_x / \omega = k$ where $k$ is an integer. 
For illustration purposes, we use normalized momentum values of $\varepsilon_x / \omega = 3$ and $\varepsilon_y / \omega = 0.01$.
Since the system is resonant and the transverse momentum value is small, Eq. \eqref{eq:probability2} prescribes that transition probability should take its maximal value of $1/2$ for $\eta=0$, except if the conditions for CDT are satisfied.
Regions where these conditions are satisfied are visible in Fig. \ref{fig:emap}a in the form of five distinct dips that exist for relatively small values of $\eta$ and given values of the parameter $A_0'/\omega$.
Looking at the specific case of $\eta = 0.1$, results show that the transition probability can be suppressed down to values of $\simeq 0.1$ by choosing appropriate values of $A_0'/\omega$ (Fig. \ref{fig:emap}b).
Interestingly, increasing the ellipticity allows one to increase the width of the probability dips which are too narrow to be useful at $\eta = 0$.
Further increasing $\eta$ shifts the dips towards smaller field strengths, and they are destroyed for large values of $\eta$.
The shift can be interpreted in terms of the AC Stark effect, whereas the destruction of the dips for a circular polarization is associated to the quasi-energy gap opening caused by the breaking of time-reversal symmetry.
This result shows the potential for controlling the response of graphene-based optoelectronic devices by selectively suppressing multiphoton resonant peaks via the applied field strength and ellipticity.
This is in analogy with the control of dynamical systems described by driven double-well potentials, such as ammonia \cite{Grifoni1998} or argon \cite{Kierig2008} molecules in a laser field.

Once again, the position of the transition probability dips can be related to the Floquet quasi-energy diagram shown in Fig. \ref{fig:emap}c for $\eta=0.1$.
The locations of the dips correspond to very close avoided crossings of the Floquet eigenstates.
For small values of the ellipticity and given the small value of the transverse momentum, their position can be estimated as the zeros of the derivative of the Bessel function, that is $A_0'/\omega = j'_{3,n}$, as shown in Appendix.
We have checked numerically that this behavior is general for integer values of $\varepsilon_x / \omega$, that is whenever the coupled laser-graphene system is tuned sufficiently close to a multiphoton resonance at small transverse momentum.

To conclude the discussion, let us mention that experiments to probe carrier dynamics in graphene could realistically be carried using laser excitations in the THz range, field strengths on the order of $10^7$ V/m \cite{Ropagnol2013,Hafez2016}, and techniques such as ARPES.
The main impediment for the realization of these experiments lies in the fact that, if electron-electron interactions can not be neglected, the carrier lifetime in pristine graphene is limited to $\sim 10$ fs \cite{Gierz2016}.
A detailed discussion of the experimental feasibility of momentum space experiments with graphene can be found in Ref. \cite{Fillion-Gourdeau2016}.

\section{Summary}
In this work, a detailed theoretical analysis of the problem of electron-hole pair creation using graphene irradiated by an elliptically polarized laser excitation was presented.
Concentrating on the CW regime, the photo-induced conduction band populations were numerically computed using Floquet theory.
We started by presenting the paradigmatic cases of linear and circular polarization, which are respectively related to Landau-Zener-St\"{u}ckelberg interferometry and the emergence of a Floquet topological insulator phase.
Most of the momentum space features can be explained in terms of the avoided crossing structure of the Floquet eigenstates.
Results for an elliptic polarization were subsequently obtained, and we showed that the ellipticity of the laser excitation provides an additional control parameter over the phenomenon of coherent destruction of tunneling in graphene.
In addition to being useful for the prediction of spectroscopy experiments, these results highlight the possibility of using ellipticity as a further optimization variable to maximize or minimize electron-hole pair production in graphene.
This optimization variable could be used in conjunction with pulse shaping \cite{Grifoni1998}, for example.

\begin{acknowledgments}
	Computations were made on the supercomputer \emph{Mammouth} from Universit\'e de Sherbrooke, managed by Calcul Québec and Compute Canada.
	The operation of this supercomputer is funded by the Canada Foundation for Innovation (CFI), minist\`ere de l'\'{E}conomie, de la Science et de l'Innovation du Qu\'{e}bec (MESI) and the Fonds de recherche du Qu\'ebec -- Nature et technologies (FRQNT).
\end{acknowledgments}

\appendix
\section{Appendix: Quasi-energy extrema}
\label{sec:appendix}
At small transverse momentum ($\varepsilon_y / \omega \rightarrow 0$) and for a linear polarization ($\eta=0$), it is possible to predict analytically the position of avoided crossings in the quasi-energies, as displayed in Fig. \ref{fig:emap}.
Starting from the perturbation approach presented by Son \emph{et al.}, one obtains the following approximate formula for quasi-energies \cite[Eqs. (26--27)]{Son2009}
\begin{equation}
q_\pm = -\frac{k \omega}{2} \pm \sqrt{\lambda J_{-k} (z) + (k \omega - \varepsilon_x + 2 \delta)^2 / 4},
\end{equation}
where $z \equiv A'_0/\omega$, $\lambda \equiv - \varepsilon_y / 2$ and $\delta$ is a level shift term of order $\lambda^2$ which also depends on $z$.
Since the $\delta$ term is small, we take the derivative of $q_\pm$ with respect to $z$ by considering only the term proportional to $\lambda$ under the square root:
\begin{equation}
\frac{dq_\pm}{dz} = \pm \frac{\lambda^2 J_k(z) J_k'(z)}{\sqrt{\lambda J_{-k} (z) + (k \omega - \varepsilon_x + 2 \delta)^2 / 4}}.
\end{equation}
From this formula, one can clearly show that the location of avoided crossings as $z$ is varied are given by the Bessel function zeros $j_{k,n}$ and $j'_{k,n}$, with the latter corresponding to the dashed lines in Fig. \ref{fig:emap} for $k=3$.

\bibliography{extracted}

\end{document}